\documentclass{article}
\pdfoutput=1
\usepackage{natbib}
\usepackage{graphicx}
\usepackage{newalg}

\title{Selective Association Rule Generation}
\author{Michael Hahsler and Christian Buchta and Kurt Hornik}

\newcommand{\set}[1]{\mathcal{#1}}

\begin{document}
\maketitle

\abstract{Mining association rules is a popular and well researched
  method for discovering interesting relations between variables in
  large databases. A practical problem is that at medium to low support
  values often a large number of frequent itemsets and an even larger
  number of association rules are found in a database.  A widely used
  approach is to gradually increase minimum support and minimum
  confidence or to filter the found rules using increasingly strict
  constraints on additional measures of interestingness until the set of
  rules found is reduced to a manageable size.  In this paper we describe
  a different approach which is based on the idea to first define a set
  of ``interesting'' itemsets (e.g., by a mixture of mining and expert
  knowledge) and then, in a second step to selectively generate rules
  for only these itemsets.  The main advantage of this approach over
  increasing thresholds or filtering rules is that the number of rules
  found is significantly reduced while at the same time it is not
  necessary to increase the support and confidence thresholds which
  might lead to missing important information in the database.}

\section{Motivation}

Mining association rules is a popular and well researched method for
discovering interesting relations between variables in large databases.
\cite{arules:Piatetsky-Shapiro:1991} describes analyzing and presenting
strong rules discovered in databases using different measures of
interestingness. Based on the concept of strong rules,
\cite{arules:Agrawal+Imielinski+Swami:1993} introduced association rules
for discovering regularities between products in large scale transaction
data recorded by point-of-sale systems in supermarkets.  For example,
the rule
\[\{\mathrm{onions, vegetables}\} \Rightarrow \{\mathrm{beef}\}\]
found in the sales data of a supermarket would indicate that if a
customer buys onions and vegetables together, he or she is likely to also
buy beef. Such information can be used as the basis for decisions about
marketing activities such as, e.g., promotional pricing or product
placements.  Today, association rules are employed in many application
areas including Web usage pattern analysis~\citep{Srivastava2000},
intrusion detection~\citep{Luo2000} and
bioinformatics~\citep{Creighton2003}.

Formally, the problem of mining association rules from transaction data
can be stated as follows~\citep{arules:Agrawal+Imielinski+Swami:1993}.
Let $I=\{i_1, i_2,\ldots,i_n\}$ be a set of $n$ binary attributes called
\emph{items}.  Let $\set{D} = \{t_1, t_2, \ldots, t_m\}$ be a set of
transactions called the \emph{database}.  Each transaction in~$\set{D}$ has a
unique transaction ID and contains a subset of the items in~$I$.
A \emph{rule} is defined as an implication of the form $X \Rightarrow Y$
where $X, Y \subseteq I$ and $X \cap Y = \emptyset$.  The sets of items
(for short \emph{itemsets}) $X$ and $Y$ are called \emph{antecedent}
(left-hand-side or LHS) and \emph{consequent} (right-hand-side or RHS)
of the rule, respectively.

To select interesting rules from the set of all possible rules,
constraints on various measures of significance and strength can be
used.  The best-known constraints are minimum thresholds on support and
confidence.  The \emph{support} of an itemset  is defined as the proportion
of transactions in the data set which contain the itemset.  All itemsets
which have a support above a user-specified minimum 
support threshold are called
\emph{frequent itemsets}.
The \emph{confidence} of a rule $X\Rightarrow Y$ is defined as
$\mathrm{conf}(X\Rightarrow Y) = \mathrm{supp}(X \cup Y) / \mathrm{supp}(X)$.
This can be interpreted as an estimate of the probability $P(Y|X)$, the
probability of finding the RHS of the rule in transactions under the condition
that these transactions also contain the LHS
\citep[e.g.,][]{arules:Hipp+Guentzer+Nakhaeizadeh:2000}.

Association rules are required to satisfy a user-specified minimum support and
a user-specified minimum confidence at the same time.  Association rule
generation is always a two-step process. First, minimum support is applied to
find all frequent itemsets in a database.  In a second step, these frequent
itemsets and the minimum confidence constraint are used to form rules.

At medium to low support values, usually a large number of frequent itemsets
and an even larger number of association rules are found in a database
which makes analyzing the rules extremely time consuming or even impossible.
Several solutions to this problem were proposed. A practical strategy is to
either increase the user-specified support or confidence threshold to reduce
the number of mined rules. It is also popular to filter or rank found rules
using additional interest measures (e.g., the measures
analyzed by \cite{arules:Tan:2004}).
However, increasing thresholds and filtering rules till a manageable number is
left can lead to the problem that only already obvious and well-known rules are
found.

Alternatively, each rule found can be matched against a set of
expert-generated rule templates to decide whether it is interesting or
not~\citep{arules:Klemettinen+Mannila+Ronkainen+Toivonen+Verkamo:1994}.
For the same purpose, \cite{Imielinski1998} describe a query language to
retrieve rules matching certain criteria from a large set of mined
rules. A more efficient approach is to apply additional constraints on
item appearance or on additional interest measures already during mining
itemsets~\citep[e.g.,][]{arules:Bayardo+Agrawal+Gunopulos:2000,
  arules:Srikant+Vu+Agrawal:1997}.  With this technique, the time to
mine large databases and the number of found itemsets can significantly
be reduced.  The popular Apriori implementation by \cite{Borgelt2006} as
well as some commercial data mining tools provide a similar mechanism
where the user can specify which items have to or cannot be part of the
LHS or the RHS of the rule.

In this paper we discuss a new approach. Instead of treating
mining association rules from transaction data as a single two-step
process where maybe the structure of rules can be specified (e.g., by
templates), we completely decouple rule generation from frequent itemset
mining in order to gain more flexibility. With our approach, 
rules can be generated from an
arbitrary sets of itemsets. This gives the analyst the possibility to
use any method to define a set of ``interesting'' itemsets and then
generate rules from only these itemsets.  Interesting itemsets can be
the result of using a mixture of additional constraints during mining,
arbitrary filtering operations and expert knowledge.

\section{Efficient selective rule generation\label{sec:generation}} 
For convenience, we introduce $\set{X} = \{X_1, X_2, \ldots, X_l\}$ for sets of
$l$ itemsets.  Analogously, we define $\set{R}$ for sets of rules.

Generating association rules is always separated into two tasks, first, mining
all frequent itemsets~$\set{X}_f$ and then generating a set of rules~$\set{R}$
from $\set{X}_f$. Extensive research exists for the first task~\citep[see,
e.g.,][]{arules:Hipp+Guentzer+Nakhaeizadeh:2000,arules:Goethals+Zaki:2004},
therefore, we concentrate in the following 
on the second task, the rule generation.

In the general case of rules with an arbitrary size of the rule's
right-hand-side, for each itemset $Z \in \set{X}$ with size $k$ we have
to check confidence for $2^k - 2$ rules $Z \setminus Y \Rightarrow Y$
resulting from using all non-empty proper subsets~$Y$ of $Z$ as a rule's
RHS.  For sets with large itemsets this clearly leads to an enormous
computational burden.  Therefore, most implementations and also this
paper follows the original definition of
\cite{arules:Agrawal+Imielinski+Swami:1993} who restrict $Y$ to single
items, which reduces the problem to only $k$ checks for an itemset of
length $k$.

The rule generation engine for the popular Apriori algorithm (e.g., the
implementation by \cite{arules:Borgelt:2003,Borgelt2006}) efficiently
generates rules by reusing the data structure built level-wise
during counting the
support and determining the frequent itemsets. The data structure
contains all support information and provides fast access for
calculating rule confidences and other measures of interestingness.

If a set of itemsets~$\set{X}$ is generated by some other means, no such data
structure might be available. Since the downward-closure property of
support~\citep{arules:Agrawal+Srikant:1994} guarantees that for a frequent
itemset also all its subsets are frequent, the data structure can be rebuilt
from a complete set of all frequent itemsets and their support values.
However, the aim of this paper is to efficiently induce rules from an arbitrary
set of itemsets which, e.g., could be specified by an expert without the help
of a data mining tool. In this case, the  support information needed to
calculate confidence is not available. For example, if all available
information is an itemset containing five items and it is desired to generate
all possible rules containing all items of this itemset, the support of the
itemset (which we might know) and the supports of all its subsets of length
four are needed. This missing support information has to be obtained from the
database.

A simple method would be to reuse an implementation of 
the Apriori algorithm with the support of the
least frequent itemset in $\set{X}$. If this support is known, $\set{X}_f
\supseteq \set{X}$ will be found.  Otherwise, the user has to iteratively
reduce the minimum support till the found $\set{X}_f$ contains all itemsets in
$\set{X}$. The rule generation engine will then produce the set of all
rules which can be generated for all itemsets in $\set{X}_f$. From this set all
rules which do not stem from the itemsets in $\set{X}$ have to be 
filtered, leaving
only the desired rules. Obviously, this is an ineffective method which
potentially generates an enormous number of rules of which the majority has to
be filtered, representing an additional large computational effort. The problem
can be reduced using several restrictions. For example, we can restrict the
maximal length of frequent itemsets to the length of the longest itemset in
$\set{X}$.  Another reduction of computational complexity can be achieved by
removing all items which do not occur in an itemset in $\set{X}$ from the
database before mining. However, this process is still far from being efficient,
especially if many itemsets in $\set{X}$ share some items or if $\set{X}$
contains some very infrequent itemsets.

To efficiently generate rules for a given confidence or other measure of rule
strength from an arbitrary set of itemsets $\set{X}$ the following
steps are necessary:

\begin{enumerate}
\item Count  the support values each itemset $X \in \set{X}$ and the
    subsets $\{X \setminus \{x\}: x \in X\}$ needed for rule generation
    in a single pass over the database
    and store them in a suitable data structure.
\item Populate set $\set{R}$ by selectively generating only rules for the
    itemsets in $\set{X}$ using the support information 
    from the data structure created in step 1.
\end{enumerate}

This approach has the advantage that no expensive rule filtering is necessary
and that combinatorial explosion due to some very infrequent itemsets
in $\set{X}$ is avoided.

\begin{figure}[tp]
\centering
\includegraphics[width=9cm]{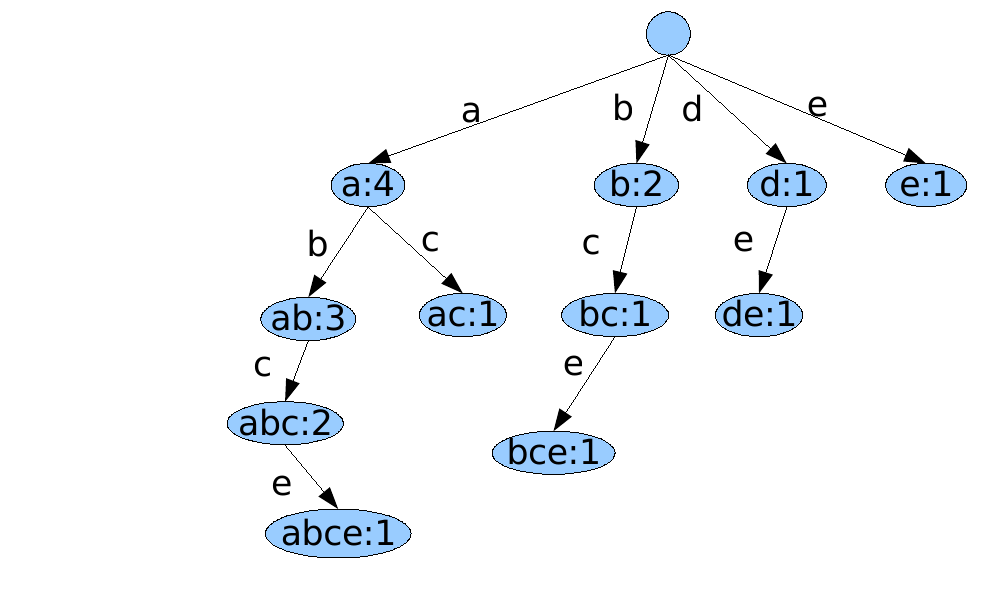} 
\caption{Prefix tree as a condensed representation of a database.
\label{fig:prefixtree}}
\end{figure}

The data structure for the needed support counters needs to provide fast
access for counting and retrieving the support counts.  A suitable data
structure is a prefix tree~\citep{Knuth1997}.  Typically, prefix trees
are used in frequent itemset mining as condensed representations for the
databases. Here the items in transactions are lexically ordered and each
node contains the occurrence counter for a prefix in the transactions.
The nodes are organized such that nodes with a common prefix are in the
same subtree.  The database in Table~\ref{tab:db} is represented by the
prefix tree in Figure~\ref{fig:prefixtree} where each node contains a
prefix and the count for the prefix in the database.  For example, the
first transaction~$\{a, b, c\}$ was responsible for creating (if the
nodes did not already exist) the nodes with the prefixes~$a$, $ab$ and
$abc$ and increasing each node's count by one.

Although adding transactions to a prefix tree is very efficient,
obtaining the counts for itemsets from the tree is expensive since
several nodes have to be visited and their counts have to be added up.
For example, to retrieve the support of itemset~$X=\{d,e\}$ from the
prefix tree in Figure~\ref{fig:prefixtree}, all nodes except $abce$,
$bce$ and $e$ have to be visited.  Therefore, for selective rule
generation, where the counts for individual itemsets have to be
obtained, using such a transaction prefix tree is not very efficient.

\begin{table}
    \centering
    \begin{tabular}{cl}
        \hline
        TID & Items\\
        \hline
        1 & $\{a, b, c\}$ \\
        2 & $\{b, c, e\}$ \\
        3 & $\{e\}$ \\
        4 & $\{a, b, c, e\}$ \\
        5 & $\{b\}$ \\
        6 & $\{a, c\}$ \\
        7 & $\{d, e\}$ \\
        8 & $\{a, b\}$ \\ 
        \hline
\end{tabular}
\caption{Example database\label{tab:db}}
\end{table}

\begin{figure}[tp]
\centering
\includegraphics[width=9cm]{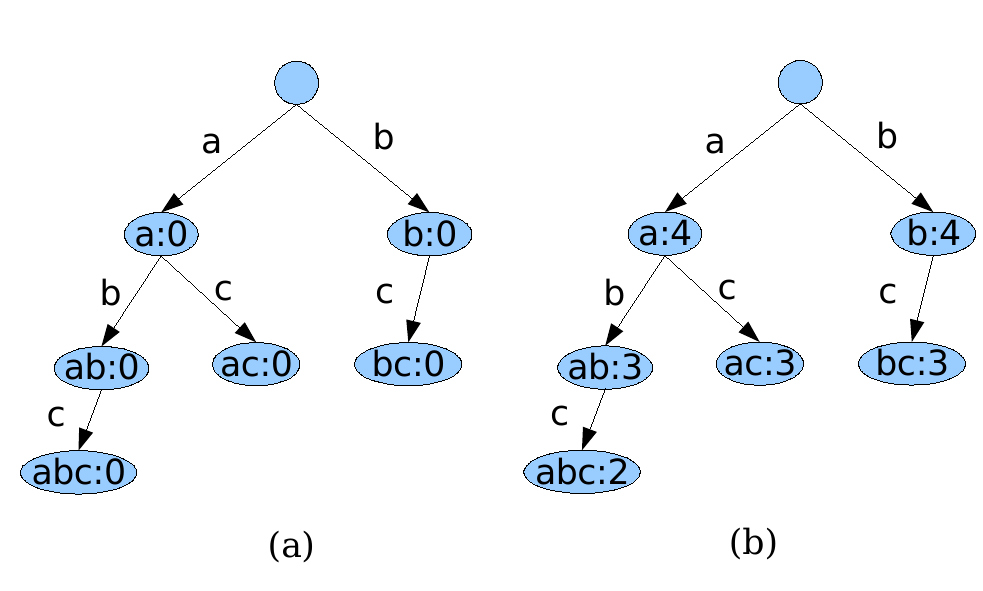} 
\caption{Prefix tree for itemset counting. (a) contains the empty 
tree to count the needed itemsets for rules containing $\{a, b, c\}$ and
(b) contains the counts.
\label{fig:itemset_prefixtree}}
\end{figure}

For selective rule generation we use a prefix tree 
similar to the itemset tree described by~\cite{arules:Borgelt+Kruse:2002}.
However, we
do not generate the tree level-wise, but we
first generate a prefix tree which only contains the nodes necessary to hold
the counters for all itemsets which need to be counted.  For example, for
generating rules for the itemset $\{a, b, c\}$, we need to count the itemset
and in addition $\{a, b\}$, $\{a, c\}$ and $\{b, c\}$. The corresponding prefix
tree is shown in Figure~\ref{fig:itemset_prefixtree}(a). The tree contains the
nodes for the itemsets plus the necessary nodes to make it a prefix tree
and all counters are initialized with zero.  Note
that with an increasing number of itemsets, the growth of nodes in the tree
will decrease since itemsets typically share items and thus will also share
nodes in the tree.

\begin{table}
\centering
\begin{minipage}{1pt}
\begin{algorithm}{count}{t,p}
\begin{IF}{t.size > 0}
   n \= successor(t[1], p) \\  
   \begin{IF}{n \ne \NIL}
      n.counter{++}   \\  
      \CALL{count}(t[2\ldots k], n)
\end{IF}\\
\CALL{count}(t[2\ldots k], p)
\end{IF}\\
\RETURN
\end{algorithm}
\end{minipage}
\caption{Recursive itemset counting function\label{tab:pseudocode}}
\end{table}



After creating the tree, we count the itemsets for each transaction
using the recursive function in
Table~\ref{tab:pseudocode}.
The function \textsc{count}$(t,p)$ is called with a transaction (as an
array $t[1\ldots k]$ representing a totally ordered set of items) and
the root node of the prefix tree.  Initially, we test if the transaction
is empty (line 1) and if so, the recursion is done.  In line 2, we try
to get the successor node of the current node that corresponds to the
first item in $t$.  If a node $n$ is found, we increase the node's
counter and continue recursively counting with the remainder of the
transaction (lines 4 and 5). Otherwise, no further counting is needed in
this branch of the tree.  Finally, we recursively count the transaction
with the first element removed also into the subtree with the root node
$p$ (line 6). This is necessary to count all itemsets that are covered
by a transaction.  For example, counting the transaction $\{a,b,c,e\}$
in the prefix tree in Figure~\ref{fig:itemset_prefixtree} increases the
nodes $a$, $ab$, $abc$, $ac$, $b$, and $bc$ by one.

There are several options to implement the structure of an $n$-ary
prefix tree (e.g., each node contains an array of pointers or a hash
table is used).  In the implementation used for the experiments in this
paper, we use a linked list to store all direct successors of a
node. This structure is simple and memory-efficient but has the price of
an increased time complexity for searching a successor node in the
recursive itemset counting function (see line 2 in
Table~\ref{tab:pseudocode}). However, this drawback can be mitigated by
first ordering the items by their inverse item-frequency. This makes
sure that items which occur often in the database are always placed near
to the beginning of the linked lists.

After counting, the support for each itemset is contained in the node
with the prefix equal to the itemset.  Therefore, we can retrieve the needed
support values from the prefix tree and generating the rules is straight
forward.

\section{Experimental results}

We implemented the proposed selective rule generation procedure using C
code and added it to the
R~package~arules~\citep{Hahsler2007}\footnote{The source code is freely
  available and can be downloaded together with the package arules from
  \url{http://CRAN.R-project.org}. Selective rule generation was added
  to arules in version 0.6-0.}.  To examine the efficiency we use the
three different data sets shown in Table~\ref{charcteristics}.  The
\emph{Adult} data set was extracted by~\cite{Kohavi1996} from the census
bureau database in 1994 and is available from the UCI Repository of
Machine Learning Databases~\citep{arules:Newman+Hettich+Blake:1998}. The
continuous attributes were mapped to ordinal attributes and each
attribute's values was coded by an item.  The recoded data set is also
included in package arules.  \emph{T10I4D100K} is an artificially
generated standard data set using the procedure presented
by~\cite{arules:Agrawal+Srikant:1994} which is used for evaluation in
many papers. \emph{POS} is an e-commerce data set containing several
years of point-of-sale data which was provided by Blue Martini Software
for the KDD Cup 2000~\citep{Kohavi2000}. The size of these three data
sets varies from relatively small with less than 50,000 transactions and
about 100 items to more than 500,000 transactions and 1,500 items. Also
the sources are diverse and, therefore, using these data sets should
provide insights into the efficiency of the proposed approach.

\begin{table}
    \centering
    \begin{tabular}{lccc} 
        \hline  
                &{\bf Adult}    & {\bf T10I4D100K} & {\bf POS}   \\ 
        \hline  
Source          &questionnaire&artificial &e-commerce            \\
Transactions    &48,842         &100,000    &515,597           \\
Mean transaction size&12.5           &10.1       &6.5               \\
Median transaction size&13.0         &10.0       &4.0               \\
Distinct items  &115            &870        &1,657             \\ 
        \hline                               
Min.~support    &0.002          &0.0001     &0.00055            \\
Min.~confidence &0.8            &0.8        &0.8                \\
Frequent itemsets &466,574        &411,365    &466,999           \\
Rules           &1.181,893      &570,908    &361,593          \\
        \hline 
    \end{tabular} 
    \caption{The used data sets.\label{charcteristics}} 
\end{table}

We compare the running time behavior of the
proposed rule generation method with the highly optimized Apriori 
implementation by \cite{Borgelt2006} which produces association rules. 
For Apriori, we use the following settings:
\begin{itemize}
\item Instead of the support stated in
    Table~\ref{charcteristics}, we use the smallest support value of an itemset
    in $\set{X}$ as the minimum support constraint and we restrict mining to
    itemsets no longer than the longest itemset in $\set{X}$.  Also, we remove
    all items which do not occur in $\set{X}$ from the database prior to
    mining.  These settings significantly reduce the search space and therefore
    also Apriori's execution time. However, it has to be noted that setting the
    minimum support requires that the support of all itemsets in $\set{X}$ is
    known. This is not the case if some itemsets in $\set{X}$ are defined by an
    expert without mining the database. In this case, one would have to use
    trial and error to find the optimal value.
\item For the comparison, we omit the expensive filter operation
    to find only the rules stemming from $\set{X}$. Therefore, 
    using Apriori for selective rule generation will take
    more than the reported time.
\end{itemize}

\begin{figure}[p]
\centering
\includegraphics[width=11cm]{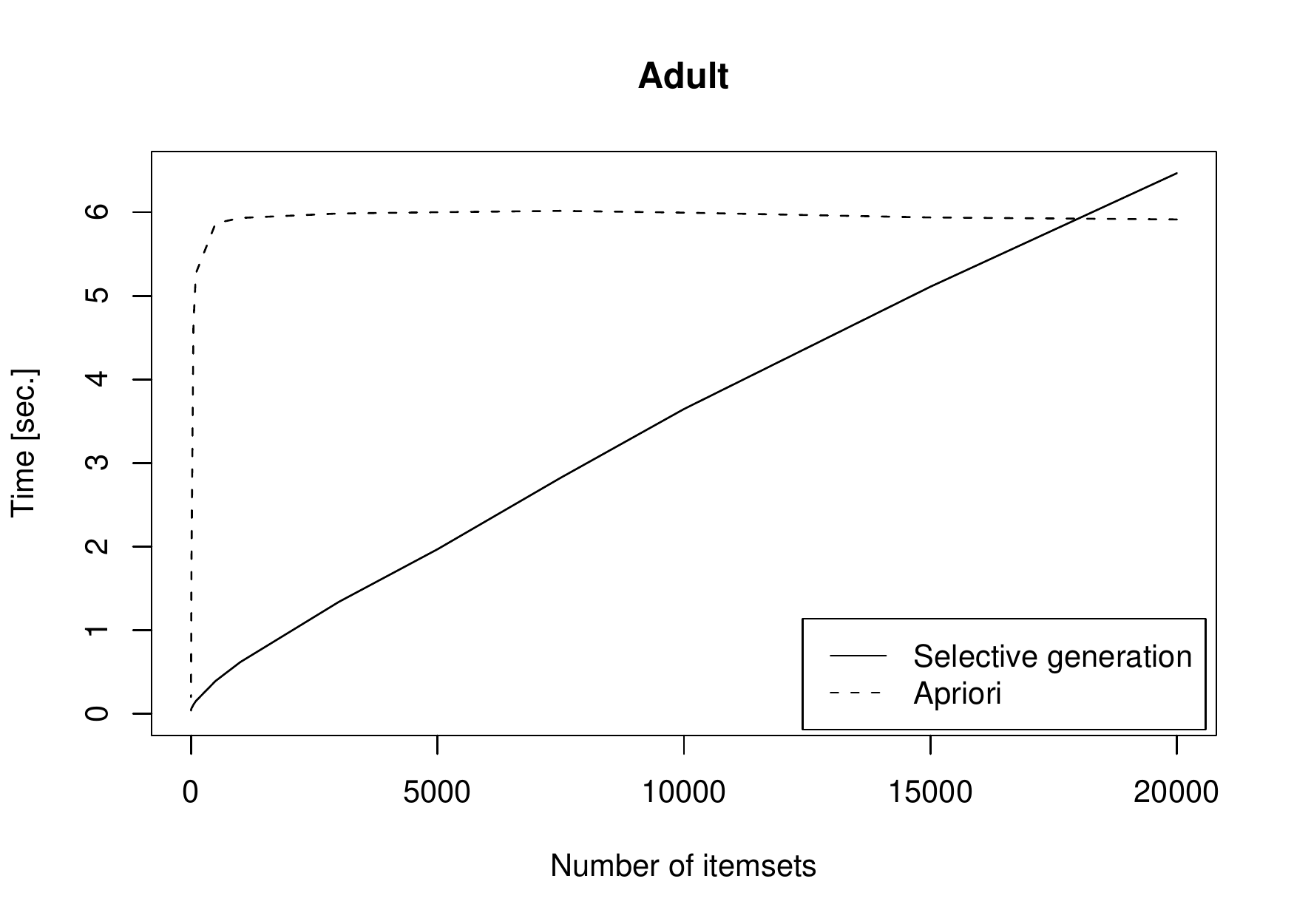} 
\caption{Running time for the Adult data set.\label{fig:Adult}}
\end{figure}

\begin{figure}[p]
\centering
\includegraphics[width=11cm]{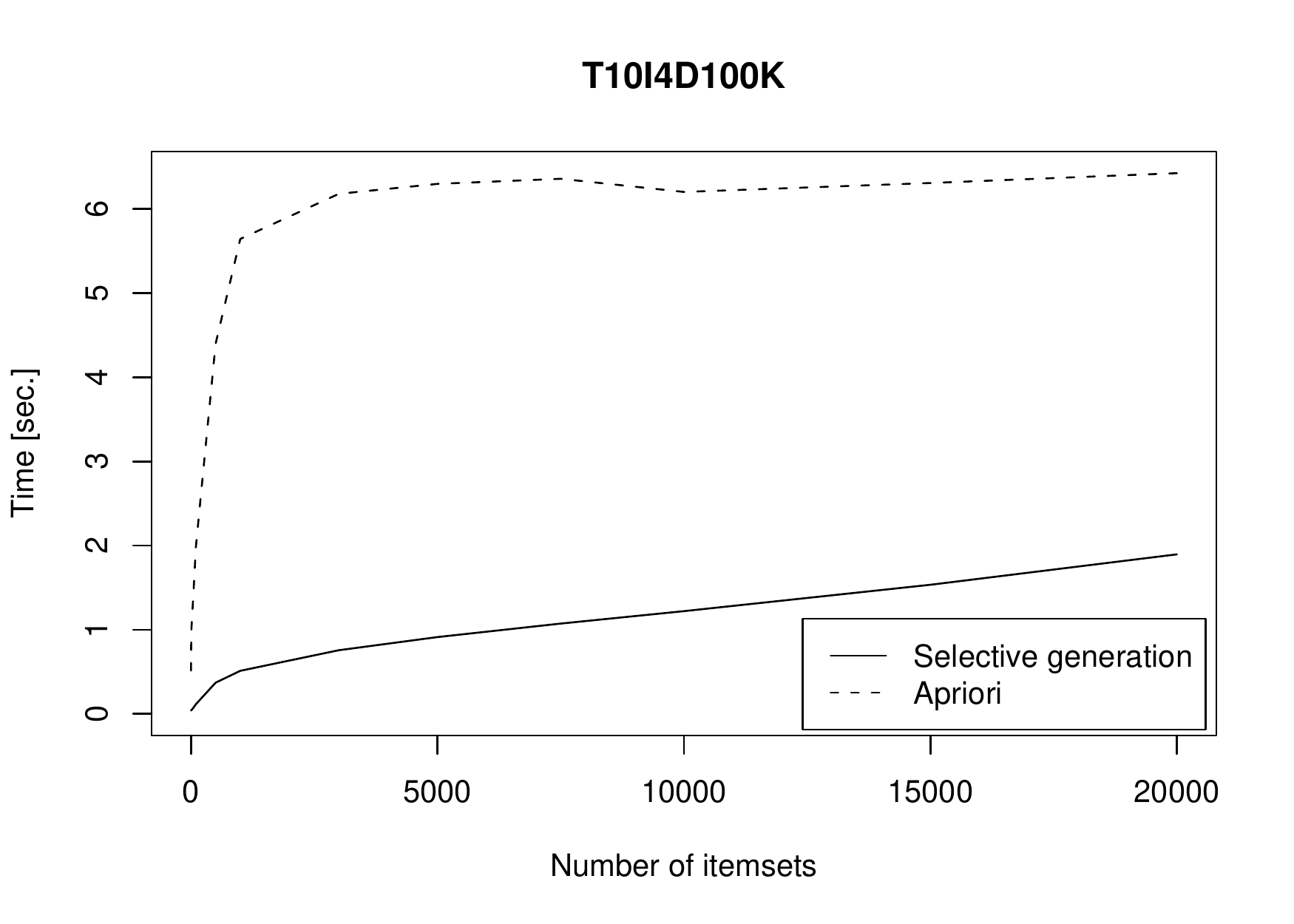} 
\caption{Running time for the T10I4D100K data set.\label{fig:T10}}
\end{figure}

\begin{figure}[tp]
\centering
\includegraphics[width=11cm]{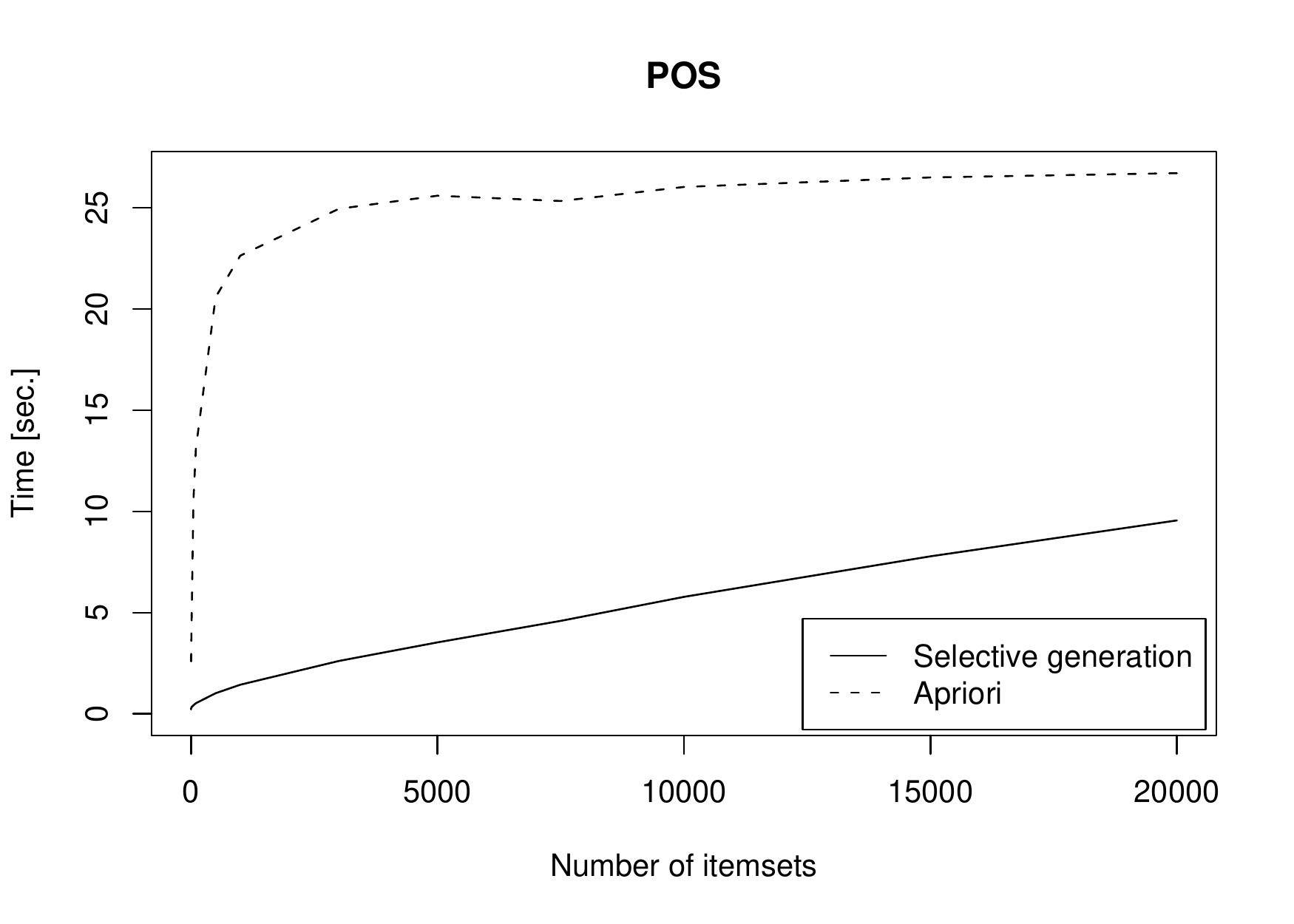} 
\caption{Running time for the POS data set.\label{fig:POS}}
\end{figure}

To generate for each data set a pool of interesting itemsets, we mine
frequent itemsets with a minimum support such that we obtain between
400,000 and 500,000 itemsets (see Table~\ref{charcteristics}). From this
pool, we take random samples which represent the sets of itemsets
$\set{X}$ we want to produce rules for.  We use a minimum confidence of
0.8 for all experiments.  We vary the size of $\set{X}$ from 1 to 20,000
itemsets, repeat the procedure for each size 100 times and report the
average execution times for the three data sets in
Figures~\ref{fig:Adult} to~\ref{fig:POS}.

For Apriori, the execution time reaches a plateau already for a few 100
to a few 1000 itemsets in $\set{X}$ and is then almost constant, for all
data sets considered.  At that point Apriori already efficiently mines
all rules up to the smallest necessary minimum support and the specified
minimum confidence.  The running time of the selective rule generation
increases sub-linearly with the number of interesting itemsets. The
increase results from the fact that with more itemsets in $\set{X}$, the
prefix tree increases in size and therefore the counting procedure has
to visit more nodes and gets slower.  The increase is sub-linear because
with an increasing number of itemsets the chances increase that several
itemsets share nodes which slows down the growth of the tree size.

Figures~\ref{fig:Adult} to~\ref{fig:POS} show that for up to 20,000
itemsets in $\set{X}$, the selective rule generation is usually much
faster than mining rules with Apriori even though the expensive
filtering procedure was omitted. Only on the Adult data set the proposed
method is slower than Apriori for more than about 18,000 itemsets in
$\set{X}$. The reason is that at some point, the prefix tree for
counting contains too many notes and performance deteriorates compared
to the efficient level-wise counting of all frequent itemsets employed
by Apriori.

The selective rule generation procedure represents an significant
improvement for selectively mining rules for a small set of (a few
thousand) interesting itemsets.  On the modern desktop PC (we used a
single core of an Intel Core2 CPU at 2.40\,GHz), the results can be
found typically under one second while using Apriori alone without
filtering takes already several times that long.  This improvement of
getting results almost instantly is crucial since it enables the analyst
to interactively examine data.

\section{Application example}

As a small example for the application of selective rule generation, we
use the Mushroom data set~\citep{arules:Newman+Hettich+Blake:1998} which
describes 23 species of gilled mushrooms in the Agaricus and Lepiota
family.  The data set contains 8124 examples described by 23
nominal-valued attributes (e.g., cap-shape, odor and class (edible or
poisonous)).  By using one binary variable for each possible attribute
value to indicate if an example possesses the attribute value, the 23
attributes are recoded into 128 binary items.

Using traditional techniques of association rule mining, an analyst
could proceed as follows.  Using a minimum support of 0.2 results in
45,397 frequent itemsets.  With a minimum confidence of 0.9 this gives
281,623 rules. If the analyst is only interested in rules which indicate
edibility, the following rule inclusive template can be used to filter
rules:

\begin{center}
    any attribute* $\Rightarrow$ any class
\end{center}

Following the notation by
\cite{arules:Klemettinen+Mannila+Ronkainen+Toivonen+Verkamo:1994}, the LHS of
the template means that it matches any combination of items for any attribute
and the RHS only matches the two items derived from the attribute class ({\tt
class=edible} and {\tt class=poisonous}).  Using the rule template to filter
the rules reduces the set to 18,328 rules which is clearly too large for visual
inspection.

For selective rule generation introduced in this paper, the expert can
decide which itemsets are of interest to gain new insights into the
data.  For example, the concept of \emph{frequent closed itemsets} can
be used to select interesting itemsets.  Using frequent closed itemsets
is an approach to reduce the number of mined itemsets without loss of
information.  An itemset is closed if no proper superset of the itemset
is contained in each transaction in which the itemset is
contained~\citep{arules:Pasquier+Bastide+Taouil+Lakhal:1999,
  arules:Zaki:2004}. Frequent closed itemsets are a subset of frequent
itemsets which preserve all support information available in frequent
itemsets.  Often the set of all frequent closed itemsets is considerably
smaller than the set of all frequent itemsets and thus easier to handle.

Mining closed frequent itemsets on the Mushroom data set with a minimum support
of 0.2 results in 1231 itemsets.  By generating rules only for these itemsets
we get 4688 rules.  Using the rule template as above leaves 154 rules, which
are way more manageable than the more than 100 times larger set obtained from
just using frequent itemsets.

\begin{table}[p]
{\small
\begin{verbatim}
  lhs                       rhs              supp. conf. 
1 {odor=none,                                                   
   gill-size=broad,                                            
   ring-number=one}      => {class=edible}  0.331    1  
2 {odor=none,                                             
   gill-size=broad,                                      
   veil-type=partial,                                    
   ring-number=one}      => {class=edible}  0.331    1  
3 {odor=none,                                             
   stalk-shape=tapering} => {class=edible}  0.307    1  
4 {odor=none,                                             
   gill-size=broad,                                      
   stalk-shape=tapering} => {class=edible}  0.307    1  
5 {odor=none,                                             
   stalk-shape=tapering,                                 
   ring-number=one}      => {class=edible}  0.307    1  

\end{verbatim}
}
\caption{Rules generated from frequent itemsets.\label{example1}}
\end{table}

\begin{table}[p]
{\small
\begin{verbatim}
  lhs                        rhs                supp. conf. 
1 {odor=none,                                                        
    gill-size=broad,                                                
    veil-type=partial,                                              
    ring-number=one}      => {class=edible}     0.331   1  
2 {odor=none,                                                 
    gill-attachment=free,                                     
    gill-size=broad,                                          
    stalk-shape=tapering,                                     
    veil-type=partial,                                        
    veil-color=white,                                         
    ring-number=one}      => {class=edible}     0.307   1  
3 {odor=none,                                                 
    gill-size=broad,                                          
    stalk-surface-below-ring=smooth,                          
    veil-type=partial,                                        
    ring-number=one}      => {class=edible}     0.284   1  
\end{verbatim}
\caption{Rules generated from frequent closed itemsets.\label{example2}}
}
\end{table}

To compare the sets of rules from the set of frequent itemsets with the
rules from the reduced set of (frequent closed) itemsets, we sort the
found rules in descending order first by confidence and then by support.
In Tables~\ref{example1} and~\ref{example2} we inspect the first few
rules of each set.  For the rules generated from frequent itemsets
(Table~\ref{example1}) we see that rules~1 and~2, and also rules~3 to~5
each have exactly the same values for support and confidence.  This can
be explained by the fact that only items are added to the LHS of the
rules which are also contained in every transaction the item in the LHS
are contained in.  For example, for rule~2 the item
\verb|veil-type=partial| is added to the LHS of rule~1.  Depending on
the type of application the rules are mined for, one of the rules is
redundant. If the aim is prediction, the shorter rule suffices.  If the
aim is to understand the structure of the data, the longer rule is
preferable.  For rules~3 to~5 the redundancy is even bigger. Inspecting
the rest of the rules
reveals that for rule~3 a total of 38 redundant rules are contained in the set.

Using closed frequent itemsets avoids such redundancies while retaining
all information which is present in the set of rules mined from frequent
itemsets.  For example, for the two redundant rules (rules~1 and 2 in
Table~\ref{example1}) the first rule with
\verb|{odor=none, gill-size=broad, ring-number=one}| in the LHS is not
present in Table~\ref{example2}.  The second rule in
Table~\ref{example2} covers rules~3 to~5 in Table~\ref{example1} plus 35
more rules (not shown in the table).

Using closed frequent itemsets is just one option. Using selective
rule generation, the expert can define arbitrary sets of interesting itemsets
to generate rules in an efficient way.

\section{Conclusion}

Mining rules not only for sets of frequent itemsets but from arbitrary sets of
possibly even relatively infrequent itemsets can be helpful to concentrate on
``interesting'' itemsets.  For this purpose, we described in this paper how to
decouple the processes of frequent itemset mining and rule generation by
proposing an procedure which obtains all needed information in a self-contained
selective rule generation process.  Since selective rule generation does not
rely on finding frequent itemsets using a minimum support threshold,
generating itemsets from itemsets with small support does not result in
combinatorial explosion.

Experiments with several data sets show that the proposed process is
efficient for small sets of interesting itemsets. Unlike existing methods 
based on frequent itemset mining, selective rule generation can support
interactive data analysis by providing almost instantly the resulting
rules. With a small application example using frequent closed itemsets
instead as the interesting itemsets, we also illustrated that selective
association rule generation can be useful for significantly reducing the
number of rules found.

\bibliographystyle{agsm}
\bibliography{arules,association_applications,misc}

\end{document}